\documentclass[twocolumn,prb,showpacs,amsmath,amssymb]{revtex4}
\usepackage{graphics,graphpap,amssymb,psfrag,subfigure,picinpar}
\usepackage{graphicx}
\usepackage{hyperref}
\begin{document}
\title{Multiple Particle Scattering in Quantum Point Contacts}
\author{Dganit Meidan and Yuval Oreg}
\date{\today}
\affiliation{Department of Condensed Matter Physics, Weizmann
Institute of Science, Rehovot, 76100, ISRAEL }
\begin{abstract}
Recent experiments performed on weakly pinched quantum point
contacts, have shown a resistance that tend to decrease at low
source drain voltage. We show that enhanced Coulomb interactions,
prompt by the presence of the point contact, may lead to
anomalously large multiple-particle scattering at finite bias
voltage. These processes tend to decrease at low voltage, and thus
may account for the observed reduction of the resistance. We
concentrate on the case of a normal point contact, and model it by
a spinfull interacting Tomonaga-Luttinger liquid, with a single
impurity, connected to non interacting leads. We find that
sufficiently strong Coulomb interactions enhance two-electron
scattering, so as these dominate the conductance. Our calculation
shows that the effective charge, probed by the shot noise of such
a system, approaches a value proportional to $e^*=2e $ at
sufficiently large backscattering current. This distinctive
hallmark may be tested experimentally. We discuss possible
applications of this model to experiments conducted on Hall bars.

\end{abstract}
 \maketitle

\newcommand{\abs}[1]{\left\lvert #1 \right\rvert}

 \emph{Introduction.}-In the past decades, one dimensional (1D)
systems have drawn extensive experimental and theoretical work.
These studies have demonstrated the profound effects of
interactions in low-dimensional systems. The theoretical model
commonly used to describe 1D systems is the Tomonaga-Luttinger
liquid (TLL)\cite{GiamarchiOxford2004}. One way to experimentally
examine the behavior of the 1D system, is to introduce
backscattering, e.g., due to confinement by a quantum point
contact (QPC). Within the framework of the TLL model, a single
impurity scatterer reduces the conductance, as the energy scale
$\Theta = \max(T,eV)$, decreases \cite{MattisPRL1974}. Here $T$
denotes the temperature, and $V$ is the bias voltage. The reduced
conductance displays the fact that reflected particles at the QPC
modulate the local electron density \cite{MatveevGlazmanPRB1993},
and enhance the reflection of other electrons. In the case of a
finite size wire, the reduction of the conductance is cutoff by
the interacting length \cite{CommentWireLength}, and does not go
to zero as $\Theta\rightarrow 0$. Instead, at low $\Theta $
corresponding to large distances, the interactions induced by the
QPC are fairly screened, and the dependence of the conductance on
$\Theta$ reflects the state of the bulk system, away from the QPC.

While some experiments support the TLL model
\cite{AuslaenderScience2002,ChangRMP2003}, recent transport
measurements appear to be inconsistent with its
predictions\cite{PepperPRB1991MarcusPRL2002,RoddaroCondMat2005,HeiblumCondMat2003}.
Within the interacting TLL, with non interacting leads, the
conductance of a system in which the basic excitations are
electrons, e.g., a normal QPC or a Hall bar at filling $ \nu =1$,
is expected to be \emph{independent} of $\Theta $ as $\Theta
\rightarrow 0 $. Similarly, for a system in the fractional Hall
regime, $ \nu=\frac{1}{3}$, the conductance is expected to
\emph{decrease} as $\Theta\rightarrow 0$. These theoretical
predictions are at odds with measurements performed on long QPC
\cite{PepperPRB1991MarcusPRL2002}, and Hall bars at integer
\cite{RoddaroCondMat2005} and fractional \cite{HeiblumCondMat2003}
fillings, where the conductance is shown to \emph{increase} as
$\Theta \rightarrow 0$.

At low $\Theta $, the state of the system away from the point
contact, where equilibration occurs, affects the conductance and
determines the allowed scattering processes. For example, if at
the place where equilibration occurs, the system is described as a
spinfull non interacting TLL, the basic excitations are electrons.
Consequently, at low energies, only electrons can be reflected or
transmitted across the constriction. In the same way, at $\nu =1/3
$, only quasi-particles can hop from one edge to the other, at low
energies. This symmetry argument does not forbid, however,
multiple particle scattering processes, in which an integer number
of particles are simultaneously scattered back.

\begin{figure}[h]
\begin{center}
\includegraphics[width = 80mm]{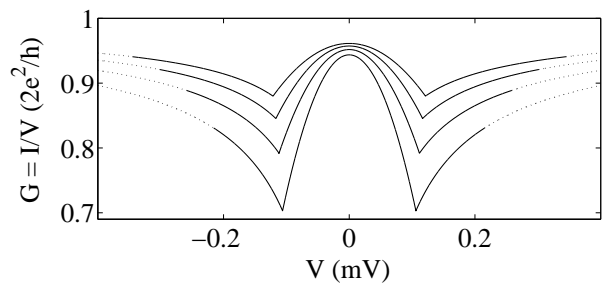}
\caption[0.4\textwidth]{ Conductance $G ={I}/{V} $ as a function
source-drain bias voltage, $V$, in the presence of single-electron
and two-electron scattering. Solid lines mark the validity of the
TLL model. Different traces are plotted at a fixed Fermi energy
($\epsilon_F\sim n^2\propto V_g^2$, where $V_g$ is the voltage on
the QPC) ranging form $5 K$ (lower curve) to $8 K $ in steps of
$1K$. We used GaAs parameters with $L=1\mu m$, $T = 80mK$, and
$\gamma\log\left({\lambda_F} /{d}\right)\approx4.3$ (see text for
definitions). \vspace{-.7cm}} \label{fig:conductance_V_sd}
\end{center}
\end{figure}

We focus our study on a normal QPC, and model it by an interacting
spinfull TLL, with non interacting leads. We will show that the
presence of strong or long ranged Coulomb interactions generate
two-electron scattering, which are relevant (in the RG sense), and
may dominate the conductance at high bias $eV
> T_L$. Here $T_L$ is the energy associated with the interacting
length~\cite{CommentWireLength,CommentGateVoltageDependence}. The
physical origin of the enhanced two-electron scattering in long
QPC's, may be deduced from extremely low densities, when the
electrons at the QPC form a periodic structure called a Wigner
crystal (WC) \cite{MatveevPRL2004,CommentMatveev}. The reciprocal
lattice spacing of the WC appears to be $K=4k_F= 2\pi n$, where $
n$ is the density, and thus it induces the backscattering of two
electrons, simultaneously. At intermediate densities a similar
density modulation appears \cite{SchulzPRL1993}, which increases
the scattering of two electrons relative to the scattering of one.
At low bias $eV<T_L$, as interactions are screened, these
processes decrease as the bias is lowered, while the single
electron process remain unchanged. As a result, the conductance is
\emph{increased} as the bias is reduced
(Fig.~\ref{fig:conductance_V_sd}), in agreement with the
experimental results \cite{PepperPRB1991MarcusPRL2002}.

\begin{figure}[h]
\begin{center}
\includegraphics[width = 80mm]{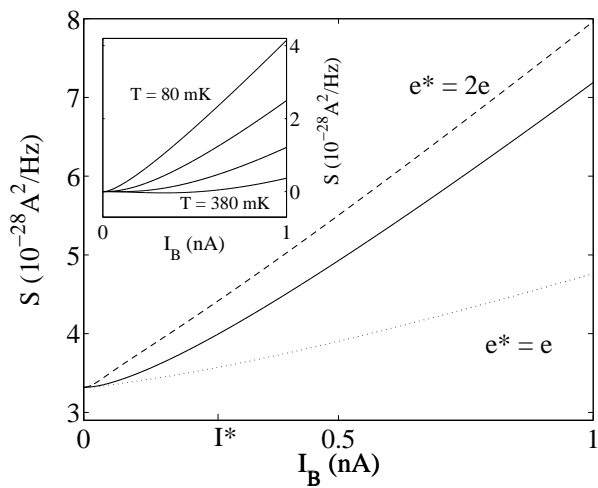}
\caption[0.4\textwidth]{ $S(\omega = 0) = \langle \delta I_T^2
\rangle_{\omega = 0}$, as a function of $I_B$ (solid line),
calculated using Ref. \onlinecite{CommentNoise}, with $\langle
\delta I_B^2\rangle_{\omega=0}$ given by Eq.~(\ref{current
correlator}),~\cite{PonomarenkoPRB1999} at $eV<T_L$. Auxiliary
plots represent the noise calculated with $\langle \delta
I_B^2\rangle_{\omega=0} = \frac{2e^*\abs{I_B}}{\tanh\left(\beta
e^* V/2 \hbar\right)}$, where $e^* = e $ (dotted line), and $e^* =
2e $ (dashed line). $I^* $ is given by E.
(\ref{backscattering_current_V<TL}), with $V=V^*$. We have used
the parameters of Fig. \ref{fig:conductance_V_sd} with $\epsilon_F
= 7 K$. Inset: Temperature dependence of the total noise, at $T $
from $ 80 mK$ to $380 mk$ in $100 mK$ steps (thermal noise was
subtracted). \vspace{-1cm} } \label{fig:shot_noise}
\end{center}
\end{figure}

We will show below that the two-electron scattering has a
distinctive signature on the shot-noise of the system. As
two-electron scattering processes dominate the conductance, the
shot-noise reveals current fluctuations with a charge that
approaches $2e $, see Fig. \ref{fig:shot_noise}. This model can be
applied to other 1D systems, e.g., a Hall bar at integer and
fractional filling, and may account for the observed conductance
behavior \cite{RoddaroCondMat2005,HeiblumCondMat2003}.

\emph{The model.}- To describe a long QPC we confine our study to
the range of gate voltage for which a single (spin degenerate)
mode crosses the constriction, yet sufficiently far from complete
depletion, such that the TLL model is applicable
\cite{CommentMatveev}. We divide the constriction created by the
QPC, into two distinct regions. The section near the QPC is
populated by a single 1D mode, and is thus characterized by
relatively low electronic density. Hence, it is modeled by an
interacting TLL. The two segments at the broadening of the
constriction are populated by a finite number of 1D modes, such
that screening is efficient. These sections are modeled by non
interacting TLL leads \cite{OregPRL1995MaslovPRB1995}. The varying
transversal width of the constriction generates a potential
modulation in the middle section, which may give rise to single
electron scattering. Since we are interested in the low energy
properties of the system, the precise form of the potential is not
essential. Therefore, we model this scattering potential as a
local impurity scatterer.

The bosonic action of the TLL with single electron scattering is
given by  $S_0+S_1 $,\cite{GiamarchiOxford2004} where
\begin{eqnarray}
\nonumber
 S_0 &=& \frac{1}{\pi}\int \frac{d\omega }{2 \pi}\left[
\frac{\abs{\omega}}{K_\rho(\omega)}\abs{\phi_\rho(\omega)}^2
+\abs{\omega}\abs{\phi_\sigma(\omega)}^2\right], \\
S_1  &=& -\lambda_1\int d\tau\epsilon_F
\cos{\sqrt{2}\phi_\rho(0,\tau)}\cos{\sqrt{2}\phi_\sigma(0,\tau)},
\label{action_scattering}
 \end{eqnarray}
here $\phi_\rho$ and $\phi_\sigma$ are related to the charge and
spin densities via
 $\rho(x) =-(\sqrt{2}/\pi)\partial_x\phi_\rho$, $\sigma_z(x)
 =-(\sqrt{2}/\pi)\partial_x\phi_\sigma$, $\lambda_1$~is the dimensionless strength of the single-electron
scattering potential, and $\epsilon_F $ is the Fermi energy (which
serves as the cutoff energy), defined as the energy difference
between the chemical potential and the bottom of the band, in the
1D section. It is therefore set by the local gate voltage. The
interaction parameter $K_\rho(\omega)$ is
\begin{eqnarray}\label{integral_eq_K_rho}
 K_\rho(\omega) &=&  2 \abs{\omega} \int\frac{dq}{2\pi}\frac{
 K_\rho(q)u_\rho(q)}{\omega^2+u_\rho(q)^2q^2},
\end{eqnarray}
with $u_\rho(q) = \sqrt{\left(v_F+g_4(q)/\pi
\right)^2-\left(g_2(q)/\pi \right)^2}$ and $K_\rho(q) =
\sqrt{\left[\pi v_F+g_4(q)-g_2(q)\right]/\left[\pi
v_F+g_4(q)+g_2(q)\right]}$. Here $g_4(q)$ and $g_2(q)$ are the
Fourier transforms of the Coulomb interaction between charge
density of the same and opposite chirality, respectively
(screening due to external metallic gates should be taken into
account). The Coulomb interaction is taken to be $V(x) =
e^2/(\epsilon_r{\sqrt{x^2+d^2}})e^{-\sqrt{x^2+d^2}/\alpha}$, where
$\alpha$ is the screening length, $\epsilon_r$ is the dielectric
constant, and the short distance singularity is cut off by the
finite width of the wire $d$. This has a Fourier transform
$V(q)\xrightarrow{qd\ll1, d\ll\alpha}
-2e^2/\epsilon_r\log{[d\sqrt{q^2+\left(1/\alpha\right)^2}]} $.
Using Eq. (\ref{integral_eq_K_rho}), with $g_2(q) = g_4(q) = V(q)$
one obtains
\begin{eqnarray}\label{K omega}
K_\rho(\omega)&\approx& 1/\sqrt{\tilde{r}_s^2
\log\left[\left(v_F\tilde{r}_s\right)/\left(d\sqrt{\omega^2 +
\omega_\texttt{SC}^2}\right)\right]},
\end{eqnarray}
where $\tilde{r}_s= \sqrt{4e^2/(\epsilon_r\pi\hbar v_F)}$, and
$\omega_\texttt{SC} = v_F \tilde{r}_s/\alpha $.

In the presence of Coulomb repulsion, a reflected electron at the
QPC may cause additional electrons to be simultaneously reflected.
An RG analysis shows that the two-electron scattering is the
principal process of such multiple-electron scattering
\cite{FurusakiPRB1993}. Long range interaction further enhance
these processes, due to the formation of a periodic density
modulation \cite{SchulzPRL1993}. We thus introduce the
two-electron scattering
\begin{eqnarray}\label{two-electron scattering lagrangian}
 \nonumber
  S_2  &=&-\lambda_2\sum_s\int d\tau ({\pi}/{k_F})^2
  \epsilon_F\\
  \nonumber
  &\times&
  \left\{R^\dag_{s}(0,\tau)R^\dag_{\bar{s}}(0,\tau)L_{\bar{s}}(0,\tau)L_{s}(0,\tau)+\textrm{h.c.}\right\}\\
 &=&-\lambda_2 \epsilon_F\int
 d\tau\cos{2\sqrt{2}\phi_\rho(0,\tau)}.
\end{eqnarray}
Using perturbation theory in the Coulomb interaction, one obtains
a lower bound assessment for the bare two-electron scattering
\cite{CommentDiagrams}
\begin{equation}\label{bare 4kf coefficient}
\lambda_2^0=\gamma
\left(\lambda_1^0\right)^2\left[{e^2}/(\epsilon_r{\hbar
    v_F})\right]\log\left({\lambda_F}/{d}\right),
\end{equation}
where $ \gamma$ is of order $1$. The factor $
\log\left(\lambda_F/d\right)$ is due to the long range nature of
the Coulomb interaction, and may be modified where short range
interactions are considered \cite{CommentBareScattering}. Eq.
(\ref{two-electron scattering lagrangian}) shows that the
two-electron scattering process involves a pair of spin up and
spin down electrons. Therefore as the degeneracy is lifted, in the
presence of a magnetic field, these scattering events will become
scarce, and the conductance resumes its spin polarized non
interacting value, $G = (1/2) G_0,\; G_0=2e^2/h$, in agreement
with the experimental observations
\cite{PepperPRB1991MarcusPRL2002}.

The scaling equations for the single-electron $\lambda_1$, and the
two-electron $\lambda_2$ scattering processes are
\begin{eqnarray}\label{scaling_eq}
\nonumber {d\lambda_1}/{d(-\log \Lambda)}&=&
\lambda_1\left[1-\left(1+K_\rho(\Lambda)\right)/2\right]\\
 {d\lambda_2}/{d(- \log{\Lambda})} &=&
\lambda_2\left[1-2K_\rho(\Lambda)\right],
\end{eqnarray}
where $\Lambda $ is the running scale. When short range Coulomb
interactions are considered, i.e., $\omega < \omega_\texttt{SC} $,
the interaction parameter $K_\rho$ is $\Lambda $ independent. Eq.
(\ref{scaling_eq}) demonstrates that in the presence of strong
short range interactions, namely for $K_\rho<\frac{1}{3} $, the
two-electron scattering renormalizes faster than the single
electron process. This tendency is enhanced in the presence of
long range Coulomb interactions, i.e. $\omega > \omega_\texttt{SC}
$, as can be seen by solving Eq.~(\ref{scaling_eq}) in the proper
limit.

\emph{Conductance.}-To find corrections to the conductance due to
one- and two-electron scattering, we calculate the backscattering
current at different energy scales. Those are the cutoff energy
$\epsilon_F=v_F/\lambda_F $, the energy associated with the
interacting length
\cite{CommentWireLength,CommentGateVoltageDependence}, $T_L=
u_\rho/L$, and the energy scale governed by the experiment~$\Theta
$. We solve the flow equations for two distinct regions. At high
energies $T_L<\Lambda<\epsilon_F$, the relevant distances are
shorter than the interacting length $ L$,\cite{CommentWireLength}
and the two scattering processes, $\lambda_1$, and $\lambda_2$
scale as Eq. (\ref{scaling_eq}), with $K_\rho(\Lambda) $ given by
Eq.~(\ref{K omega}). For $T_L<\Theta$, this behavior is carried
down to $\Lambda\sim \Theta$. However, if $\Theta < T_L$ there are
energy scales for which the relevant distances are longer than
$L$, $\Theta<\Lambda<T_L$. At these energies the interactions are
absent, $K_\rho=1$, and the single-electron scattering becomes
marginal, while the two-electron scattering is irrelevant [see
Eq.~(\ref{scaling_eq})].

At $\Theta<T_L$, the backscattering current for both short and
long range interactions is
\begin{eqnarray}\label{backscattering_current_V<TL}
 \langle I_{\textrm{B}}(t) \rangle  &=& G_1V  +G_2V \left[\left({\pi T}/{T_L}\right)^{2}
+\left({eV}/{T_L}\right)^2\right],
\end{eqnarray}
where $G_1\propto G_0 \lambda_1^2$ and $G_2\propto G_0
\lambda_2^2$, and $\lambda_1 $ and $\lambda_2$ are the
renormalized parameters for one- and two-electron scattering,
respectively. For short range interaction $\lambda_1$ and
$\lambda_2$ are found using Eqs. (\ref{scaling_eq}) and (\ref{K
omega}), with $\omega<\omega_\texttt{SC} \approx~\epsilon_F$
\begin{eqnarray}\label{lambda1 lambda2 short range}
  \lambda_1 =
  \lambda_1^0{\left({T_L}/{\epsilon_F}\right)}^{\frac{K_\rho-1}{2}};
  \lambda_2 =
 \lambda_2^0{\left({T_L}/{\epsilon_F}\right)}^{2K_\rho-1}.
\end{eqnarray}
For long range interactions $\lambda_1$ and $\lambda_2$ are found
using Eqs.~(\ref{scaling_eq}) and~(\ref{K omega}), with
$\omega>\omega_\texttt{SC} \approx T_L$.
\begin{eqnarray}\label{lambda1 lambda2 long range}
\nonumber
 \lambda_1 &=&
\lambda_1^0\left({T_L}/{\epsilon_F}\right)^{-\frac{1}{2}}
e^{-\frac{1}{\tilde{r}_s^2}\left[{K_\rho(T_L)^{-1}}-{K_\rho(\epsilon_F)^{-1}}\right]}\\
\lambda_2 &=& \lambda_2^0\left({T_L}/{\epsilon_F}\right)^{-1}
e^{-\frac{4}{\tilde{r}_s^2}\left[{K_\rho(T_L)^{-1}}-{K_\rho(\epsilon_F)^{-1}}\right]}.
\end{eqnarray}
Notice that Eqs. (\ref{lambda1 lambda2 short range}) and
(\ref{lambda1 lambda2 long range}) correspond to the two extreme
limits $\omega_\texttt{SC}\approx \epsilon_F $ and
$\omega_\texttt{SC}\approx T_L $, respectively. The intermediate
case $T_L<\omega_\texttt{SC}<\epsilon_F $ (not considered here
\cite{CommentWireLength}), interpolates between the two.

When short range interactions are considered, one can obtain an
analytic expression for both $\Theta>T_L$ and $\Theta~<~T_L$.
Following Ref.~\onlinecite{MahanNewYork2000MartinCondMat2005}, the
backscattering current can be shown to have two independent
contributions $\langle I_{\textrm{B}}(t) \rangle= \langle
I_{\textrm{B}_1}(t)
 \rangle+\langle I_{\textrm{B}_2}(t)
 \rangle$, where (with $\beta = 1/k_BT $)\cite{CommentCrossTerms}
\begin{eqnarray}\label{backscattering_current_V>TL}
\nonumber
 \langle I_{\textrm{B}_1}(t) \rangle &=& G_0 V \frac{1}{4}(\lambda_1^0)^2
 f_{K_\rho+1}\left(\hbar/(\beta\epsilon_F), \beta eV/2\hbar\right)\\
\nonumber
 \langle I_{\textrm{B}_2}(t) \rangle&=& G_0 V(\lambda_2^0)^2f_{4 K_\rho}\left(\hbar/(\beta\epsilon_F),\beta
 eV/\hbar\right),\\
f_{\nu}(x,y) &=&\frac{2^{\nu-1}\pi^\nu x^{\nu-1}}{y
\Gamma(\nu)}\sinh(y)\abs{\Gamma\left(\frac{\nu}{2}+\frac{i
y}{\pi}\right)}^2.
\end{eqnarray}

In Fig. \ref{fig:conductance_V_sd} we present the conductance $G =
I/V $ in the presence of both one- and two-electron scattering,
and short range Coulomb interactions as a function of bias
voltage, for different cutoff energies. The conductance is shown
to increase as the bias voltage decreases.
The kinks in Fig. \ref{fig:conductance_V_sd} are due to the abrupt
change in the parameter $K_\rho$, taken to be $K_\rho <1 $ in the
vicinity of the QPC, and $K_\rho=1 $ at the broadening of the
constriction. In an experimental realization, the transition
between the two is smooth, and the kinks will be softened. A study
of Eq. (\ref{backscattering_current_V>TL}) shows that for $T\ll
eV$ and $K_\rho <1/4$, the contribution of the two electron
scattering to the \emph{differential} conductance may change sign.
As a result the \emph{differential} conductance at $eV>T_L$ may
exceed the universal value $2e^2/h $, and the curvature at the
high bias regime changes sign (not shown). From Eq.
(\ref{backscattering_current_V<TL}) it follows that at $\Theta \ll
T_L$, the backscattering current is dominated by single-electron
scattering, $\langle I_{\textrm{B}}(t) \rangle\approx \langle
I_{\textrm{B}_1}(t) \rangle $. Conversely at $\Theta\lesssim T_L$,
for an interacting system that exhibits an enhanced two-electron
scattering, the current is given by $\langle I_{\textrm{B}}(t)
\rangle\approx \langle I_{\textrm{B}_2}(t) \rangle $. At $T<eV $,
the crossover voltage $V^*$ can be found by equating the two
contributions to the backscattering current in
Eq.~(\ref{backscattering_current_V<TL}), and is given by $eV^*
=\sqrt{T_L^2 G_1/G_2 -\pi^2 T^2} $.

 \emph{Noise.}-The occurrence of multi-particle scattering is
accompanied by a distinctive signature on the shot noise of the
system. Shot noise measures the size of charge-transfer events. As
multi-particle scattering become frequent, the events that
determine the shot noise consist of a charge which is larger than
the single particle charge.

The expression for the symmetrized backscattering current correlator
is found to be \cite{CommentNoise,PonomarenkoPRB1999}
\begin{eqnarray}\label{current correlator}
 \langle \delta I_B^2\rangle_{\omega = 0} &=& \frac{2e\abs{\langle
I_{\textrm{B}_1}\rangle}}{\tanh\left({\beta e V}/{2 \hbar}\right)}
+\frac{4e\abs{\langle I_{\textrm{B}_2}\rangle}}{\tanh\left(\beta e
V / \hbar\right)}.
\end{eqnarray}
The noise behavior, depicted in Fig. \ref{fig:shot_noise}, is
affected by the competition between the single and two electron
scattering events. At small $V\ll V^*$ corresponding to small
backscattering current, two-electron scattering are scarce. Thus,
the charge defined as the ratio $S/(2\abs{\langle I_B\rangle} )$
is $e^* = e $. At $V> V^*$, corresponding to large backscattering
current, two-electron scattering dominant the conductance, and the
charge approaches $ e^*\approx 2e$.

 While other existing models, e.g., the Kondo model, can account for
increasing conductance at small SD bias \cite{MeirPRL2002}, the
doubling of the charge is a unique feature of the model presented
here. Moreover, the two models have different ranges of validity.
The use of the Kondo model is justified at very low densities,
where a single electron is believed to be localized at the point
contact. Conversely the model presented here is valid at densities
high enough to consider the TLL model. Thus, noise measurements
performed at different ranges of parameters, can be used to
distinguish the two behaviors.

\emph{Conclusion.}- We have studied the spinfull TLL model with
one- and two-electron scattering. Under proper conditions, the two
electron scattering processes dominate the conductance of the
system. We predict that at this limit, the shot noise will
approach $e^* = 2e$, and suggest noise measurements as a way to
substantiate the validity of our model.

The interacting TLL with a single impurity scatterer, describes a
large class of systems at the low energy limit. An example of
which is the Hall bar contracted by means of a QPC. Within the TLL
model, the backscattering of electrons at $\nu = 1 $, and of quasi
particles at $\nu = \frac{1}{3} $ cannot account for the measured
enhancement in the conductance at low bias voltage
\cite{HeiblumCondMat2003,RoddaroCondMat2005}. Conversely, multiple
particle backscattering prompt by the reduced density near the
QPC, become scarce as the bias is lowered, leading to an enhanced
conductance, concurring with the transport measurements. In a
system dominated by $n$-particle scattering at high energies, we
expect that the shot noise, at large backscattering current, will
approach $e^*=n \nu e$. For example, at $ \nu = 1/3$, an enhanced
three quasi-particle scattering, will result in an effective
charge, that approaches $e^* = 3  \nu e = e $, at large
backscattering current. The challenging task remains to find a
microscopic model that predicts such an enhanced multiple particle
scattering in the quantum Hall system.

\emph {Acknowledgements:} We would like to thank
A.~M.~Finkel'stein, Y.~Gefen, L. Goren, B.~I.~Halperin,
M.~Heiblum, Y.~Meir, N.~Ofek, R.~de~Picciotto and A.~Stern. The
study was supported by Minerva and by an ISF grant 845/04.

\vspace{-.5cm}

\bibliographystyle{amsplain}

\begin{thebibliography}{1}
\bibitem{CommentWireLength}
When short range interactions are considered, i.e., when the
screening length is much smaller than the length of the QPC, the
interacting length, $L$, is determined by the latter. In the
opposite limit, interactions are long-ranged, and $L$ is governed
by the screening length. For simplicity we do not consider the
crossover between the two.

$L$ determines the range of energies over which the two scattering
processes grow. Thus, When $L $ is longer, the scattering is
enhanced, and the conductance is reduced.
\bibitem{CommentGateVoltageDependence}
Within our model $T_L= u_\rho/L\sim v_F/L( \sqrt{1+ 2 g/\pi
v_F})$, where $v_F=\hbar k_F/m\sim n\propto V_g$, $n$ is the
density in the 1D section, and $V_g $ is the gate voltage on the
QPC. When the QPC is opened, $v_F$ increases and hence $T_L$
increases. In reality, the dependence on $V_g$ is much more
complex. However it is clear that as $n$ increases, $L$ decreases
due to efficient screening. These corrections will further
increase $T_L$ as the QPC is opened.
\bibitem{CommentDiagrams}
Interactions generate two-electron backscattering that can not be
considered as two independent single electron processes. In
diagramatic formulation, this amounts to all connected diagrams
with four external legs (two incoming right movers and two
outgoing left movers, and their hermitian conjugates). A first
order calculation of these diagrams yields Eq. (\ref{bare 4kf
coefficient}).
\bibitem{CommentBareScattering}
It should be noted that in the densities considered here,
$\left[e^2/(\hbar v_F)\right]\log\left(\lambda_F/d\right)$ is not a
small parameter. However, higher order terms are expected to further
increase the scattering coefficient, as seen from the RG analysis.
\bibitem{CommentCrossTerms}
Dimensional counting shows that cross terms
$\propto\left(\lambda_1^0 \right)^2\lambda_2^0$ are of order of
$\Theta^2$. These contradict the Fermi liquid theory (as they
imply that the leading correction to the self energy is $\sim
{\Theta} $). An explicit calculation shows that they vanish.
\end{thebibliography}

\newcommand{\noopsort}[1]{} \newcommand{\printfirst}[2]{#1}
\newcommand{\singleletter}[1]{#1} \newcommand{\switchargs}[2]{#2#1}
\providecommand{\bysame}{\leavevmode\hbox
to3em{\hrulefill}\thinspace}
\providecommand{\MR}{\relax\ifhmode\unskip\space\fi MR }
\providecommand{\MRhref}[2]{%
  \href{http://www.ams.org/mathscinet-getitem?mr=#1}{#2}
} \providecommand{\href}[2]{#2}

\end{document}